\input psfig
\documentstyle[twoside,fleqn,espcrc2]{article}

\def\mev{\,{\rm Me\kern-0.1em V}}
\def\gev{\,{\rm Ge\kern-0.1em V}}

\newcommand{\AmS}{{\protect\the\textfont2
  A\kern-.1667em\lower.5ex\hbox{M}\kern-.125emS}}

\begin{document}
\title{\Large{\bf Electromagnetic Splittings and Light Quark Masses}}
\date{\today}
\author{A.~Duncan\address{Dept. of Physics and Astronomy, 
        Univ. of Pittsburgh, Pittsburgh, PA 15260},%
        ~E.~Eichten\address{Fermilab, P.O. Box 500, Batavia, IL 60510}
        \thanks{Presenter} and%
        ~H.~Thacker\address{Dept.of Physics, University of Virginia, 
        Charlottesville, VA 22901}}
       
\begin{abstract}
{A method for computing electromagnetic properties of hadrons 
in lattice QCD is described.
The electromagnetic field is introduced dynamically, using
a noncompact formulation.  Employing enhanced electric charges, 
the dependence of the pseudoscalar meson mass on 
the (anti)quark charges and masses can be accurately calculated.
At $\beta=5.7$ with Wilson action,
the $\pi^+-\pi^0$ splitting is found to be $4.9(3)$ MeV. 
Using the measured $K^0-K^+$ splitting, we also find 
$m_u/m_d = .512(6)$.  Systematic errors are discussed. 
Preliminary results for vector meson splittings are also presented.}
\end{abstract}

\maketitle


\section{Light Quark Masses}

If a fundamental theory of quark masses ever emerges, it may be as important
to resolve the theoretical uncertainty in the light quark masses 
as it is to accurately measure the top quark mass (e.g. in deciding
whether nature avoids the strong CP problem via a massless up quark).
The particle data tables \cite{pdata}
give wide ranges for the up ($2 < m_u <8$ MeV) and 
down ($5 < m_d <15$ MeV) quarks,
while lowest order chiral perturbation theory 
\cite{leutrev} gives $m_u/m_d=0.57\pm 0.04$.
Numerical lattice calculations provide, in principle, 
a very precise way of studying the dependence
of hadron masses on the lagrangian quark mass parameters\cite{quark_mass}. 
As the electromagnetic contribution to hadronic mass splittings  
within isomultiplets is comparable to the up-down quark mass difference,
an accurate determination of the light quark masses 
requires inclusion of electromagnetic effects 
dressed by nonperturbative QCD dynamics. 
Here, we discuss a method for studying such electromagnetic 
effects\cite{emlatt}.
In addition to the SU(3) color gauge field, we introduce a U(1) 
electromagnetic field on the lattice which is treated by quenched 
Monte Carlo methods. The resulting SU(3)$\times$U(1)
(Coulomb-gauge) configurations are then analyzed by standard 
hadron propagator techniques. 
  
The small size of electromagnetic mass splittings makes their
accurate determination by conventional lattice techniques 
difficult if the electromagnetic coupling is taken at its physical value. 
We have found that calculations done at larger values 
(roughly 2 to 6 times physical) of the quark electric charges 
lead to accurately measurable isosplittings in the light pseudoscalar meson
spectrum, while still allowing  perturbative extrapolation 
to physical values. 

We proceed as follows: quark propagators are generated in the
presence of background SU(3)$\times$U(1) fields where 
the SU(3) component represents the usual gluonic
gauge degrees of freedom, while the U(1) component incorporates an abelian
photon field (with a noncompact gauge action) 
which interacts with quarks of specified electric charge. 
Quark propagators are calculated for a variety of electric charges
and light quark mass values. 
The gauge configurations were generated at $\beta=5.7$ on a
$12^3\times 24$ lattice. 200 configurations each separated by 1000 Monte
Carlo sweeps were used. In the results reported here, 
we have used four different values of charge 
given by $e_q =$0, -0.4, +0.8, and -1.2 in units in which the electron charge
is $e=\sqrt{4\pi/137} =.3028\ldots\;$. 
For each quark charge we calculate propagators for
three light quark mass values in order to allow a chiral extrapolation. 
From the resulting 12 quark propagators, 144 
quark-antiquark combinations can be formed, 
leading 78 independent meson propagators and masses.

\section{Expected Chiral Behaviour}

Once the full set of meson masses is computed, 
the analysis proceeds by a combination of chiral and QED perturbation theory. 
In pure QCD it is known  that, in the range of masses considered here, 
the square of the pseudoscalar
meson mass is accurately fit by a linear function 
of the quark masses\cite{etaprime,Xlogs}. 
We have found that this linearity persists even in the presence 
of electromagnetism\cite{emlatt}. For each of
the charge combinations studied, the dependence of the squared meson mass 
on the bare quark mass is well described by lowest order chiral 
perturbation theory. Thus we write the pseudoscalar mass squared as
\begin{equation}
\label{eq:ChPT}
m_{P}^2 = A(e_q, e_{\bar{q}}) +  m_qB(e_q,e_{\bar{q}}) +m_{\bar{q}}B(e_{\bar{q}},e_q)
\end{equation}
where $e_q, e_{\bar{q}}$ are the quark and antiquark charges, 
and $m_q, m_{\bar{q}}$ are the bare quark masses, 
defined in terms of the Wilson hopping parameter by
$(\kappa^{-1} - \kappa^{-1}_c)/2a$. (Here $a$ is the lattice spacing.)
Because of the electromagnetic self-energy shift, 
the value of the critical hopping parameter must be determined
independently for each quark charge. This is done by requiring 
that the mass of the neutral pseudoscalar meson vanish at 
$\kappa=\kappa_c$, as discussed below.

For the physical values of the quark charges, we expect that an expansion
of the coefficients $A$ and $B$ in (\ref{eq:ChPT})
to first order in $e^2$ should be quite accurate. For the larger values
of QED coupling that we use in our numerical investigation, 
the accuracy of first order perturbation theory is less clear: in fact, 
a good fit to all our data
requires small but nonzero terms of order $e^4$, corresponding to 
two-photon diagrams. Comparison of the order $e^4$ terms with those of 
order $e^2$ provides a quantitative
check on the accuracy of QED perturbation theory. 
Only those $e^4$ terms which significantly reduce the 
$\chi^2$ per degree of freedom have been kept. 

According to Dashen's theorem, in the chiral limit the value 
of $m_P^2$ is proportional to the square of the total charge.
Thus, we have also allowed the values of the critical hopping parameters for
each of the quark charges to be fit parameters, requiring 
that the mass of the neutral mesons vanish in the chiral limit. 
Thus A takes the form  
$A^{(1)}(e_q+e_{\bar{q}})^2$ to order $e^2$.
(Order $e^4$ terms here were not found necessary to fit the data.)
The coefficient $B$ in (\ref{eq:ChPT}) which parametrizes the slope of 
$m_P^2$ may also be expanded in perturbation theory. 
Of the five possible $e^4$ terms in $B^{(2)}(e_q,e_{\bar q})$, only the 
$e_q^4, e_q^3 e_{\bar q}$ and $e_q^2 e_{\bar q}^2$ terms
were found to improve the $\chi^2$. The coefficients in A and B,
along with the four values of $\kappa_c$ for the four quark charges, 
constitute a 12-parameter fit to the meson mass values. 

\section{Lattice Formulation Including EM}

We have chosen a noncompact  abelian gauge action  $S_{\rm em}$ 
to ensure that the theory is  free  in the absence of
fermions, and is always in the nonconfining, massless phase.
(Of course, lattice gauge invariance still requires a 
compact gauge-fermion coupling).
An important aspect of a noncompact formalism is the necessity 
for a gauge choice. We use QCD lattice configurations which 
have all been converted
to Coulomb gauge for previous studies of heavy-light mesons. Coulomb 
gauge turns out to be both practically
and conceptually convenient in the QED sector as well.

For the electromagnetic action, we take
\begin{equation}
\label{eq:qedact}
S_{\rm em}=\frac{1}{4e^2}\sum_{n\mu\nu}(\nabla_{\mu}A_{n\nu}-\nabla_{\nu}
A_{n\mu})^{2}
\end{equation}
with $e$ the bare electric coupling, $n$ specifies a lattice site, 
$\nabla_{\mu}$ the discrete lattice right-gradient in the $\mu$
direction and 
$A_{n\mu}$ takes on values between $-\infty$ and $+\infty$. Electromagnetic
configurations were generated using (\ref{eq:qedact}) as a Boltzmann weight,
subject to the linear Coulomb constraint $ \bar{\nabla}_{i}A_{ni}=0 $
with $\bar{\nabla}$ a lattice left-gradient operator. 
The action is Gaussian-distributed so it is a trivial matter to 
generate a completely independent set in momentum space, 
recovering the real space Coulomb-gauge configuration 
by Fast Fourier transform.
We fix the global gauge freedom remaining after
the Coulomb gauge condition is imposed 
by setting the $p=0$ mode equal to zero for the transverse modes, and the
$\vec{p}=0$ mode to  zero for the Coulomb modes on each time-slice. 
(This implies a specific treatment of finite volume effects 
which will be discussed below.)
The resulting Coulomb gauge field
$A_{n\mu}$ is then promoted to a compact link 
variable $U^{\rm em}_{n\mu}=e^{\pm iqA_{n\mu}}$
coupled to the quark field in order to describe a quark of 
electric charge $\pm qe$.  Quark propagators are then computed in the combined
SU(3)$\times$U(1) gauge field.


\section{Preliminary Results}

For charge zero quarks, propagators were calculated at 
hopping parameter 0.161, 0.165, and 0.1667, corresponding to bare quark 
masses of 175, 83, and 53 MeV
respectively. The gauge configurations are generated at $\beta = 5.7$, and 
we have taken the lattice spacing to be $a^{-1} = 1.15$ GeV.
After shifting by the improved perturbative values, we select the
same three hopping parameters for the nonzero charge quarks\cite{emlatt}. 
Because this shift turns out to be very close to the observed shift 
of $\kappa_c$, the quark masses for nonzero charge are 
nearly the same as those for zero charge.
For all charge combinations, meson masses were 
extracted by a two-exponential fit (using both smeared and local sources)
to the pseudoscalar
propagator over the time range $t = 3$ to 11. 
Errors on each mass value are obtained by a single-elimination jackknife. 
The resulting data is fitted
to the chiral/QED perturbative formula (\ref{eq:ChPT}) by $\chi^2$ minimization.
The fitted parameters are given in Table 1. 
\begin{table}
\begin{center}
\caption[tbl:ChPT]{Coefficients of fitting function, Eq.(\ref{eq:ChPT}).
Terms consistent with zero were dropped from this fit. 
Numerical values are in ${\rm GeV}^2$ and GeV for
A and B terms respectively.}
\begin{tabular}{ll}
  & Fit \\ \hline
 $A  $ & $ 0.0143(10) (e_q + e_{\bar q})^2 $ \\ 
 $B^{(0)}  $ & $ 1.594(11) $ \\
 $B^{(1)}  $ & $ 0.205(22) e_q^2 + 0.071(9) e_q e_{\bar q} 
+ 0.050(7) e_{\bar q}^2 $ \\
 $B^{(2)}  $ & $ 0.064(17) e_q^4 - 0.031(4) e_q^2 e_{\bar q}^2 $ \\
       & $ + 0.033(6) e_q^3 e_{\bar q} - 0.031(4) e_q^2 e_{\bar q}^2 $ \\
\end{tabular}
\end{center}
\end{table} 
Errors were obtained by performing the fit on each jackknifed subensemble.  
Aside from very small corrections of order $(m_d-m_u)^2$, 
the $\pi^+-\pi^0$ mass splitting is of purely electromagnetic origin,
and thus should be directly calculable by our method. Because we have
used the quenched approximation,  $u\bar{u}$ and $d\bar{d}$ mesons
do not mix. The  squared neutral
pion mass is obtained by averaging the squared masses of the 
$u\bar{u}$ and $d\bar{d}$ states. Thus, to zeroth order in $e^2$, the terms
proportional to quark mass cancel in the difference
$m_{\pi^+}^2-m_{\pi^0}^2$. 
This difference is essentially  given by the single term  
\begin{equation}
m_{\pi^+}^2-m_{\pi^0}^2 \approx A^{(1)}e^2
\end{equation}

Using the coefficients listed in Table 1, and the
experimental values of the $\pi^0, K^0,$ and $K^+$ masses,
we may directly solve the resulting three equations for the up,
down, and strange masses. The $\pi^+-\pi^0$ splitting may then be
calculated, including the very small contributions from the 
order $e^2m_q$ terms. We obtain 
\begin{equation}
m_{\pi^+}-m_{\pi^0}= 4.9\pm 0.3 {\rm MeV}
\end{equation}
compared to the experimental value of $4.6$ MeV. (The electromagnetic
contribution to this splitting is estimated \cite{Gasser} to be $4.43\pm0.03$
MeV.)
Our calculation can be compared to the value $4.4$ MeV (for $\Lambda_{\rm QCD}
= 0.3 $ GeV and $m_s = 120$ MeV) obtained
by Bardeen, Bijnens and Gerard\cite{1overN} using large N methods.
The values obtained for the bare quark masses are
\begin{equation}
m_u = 3.86(3),\;\;\; m_d= 7.54(5),\;\;\; m_s = 147(1)
\end{equation}
The errors quoted are statistical only, and are computed by a standard
jackknife procedure. The small statistical errors reflect the
accuracy of the pseudoscalar mass determinations, and should facilitate
the future study of systematic errors (primarily finite volume,
continuum extrapolation and quark loop effects)\cite{quark_mass}, 
which are expected to be considerably larger. 
The relationship between lattice bare quark masses and the familiar 
current quark masses in the $\overline{MS}$ continuum regularization 
is perturbatively calculable\cite{quark_mass}. 
For mass ratios (which are independent of renormalization prescription)
we obtain
\begin{equation}
\frac{m_d-m_u}{m_s}=.0249(3)\;,\;\;\;\frac{m_u}{m_d}=.512(6)
\end{equation}

\section{Finite Volume Corrections}

The presence of massless, unconfined degrees of
freedom implies that  finite
volume  effects are potentially  much
larger than in pure QCD, falling  as inverse powers of
the lattice size, instead of exponentially.
We have estimated the size of these  corrections phenomenologically
along the lines of Bardeen, et.al\cite{1overN}, who
model the low-$q^2$ contribution to the $\pi^+-\pi^0$ splitting
in terms of $\pi, \rho $, and $A1$ intermediate states. This analysis gives
\begin{equation}
\label{eq:Bardeen}
\delta m_{\pi}^2 = \frac{3e^2}{16\pi^2}\int_0^{M^2}
                  \frac{m_A^2 m_{\rho}^2}{(q^2+m_{\rho}^2)(q^2+m_A^2)}dq^2
\end{equation}
 We may use this result to estimate the finite volume correction by casting
 the expression as
a four-dimensional integral over $d^4q$ and
then constructing a finite volume version of it by
replacing the integrals with discrete sums (excluding
the $q=0$ mode).
For a $12^3\times 24$ box with $a^{-1}=1.15$ GeV, we find
that the infinite volume value of $5.1$ MeV is changed to 
$\delta m_{\pi}=4.8$ MeV, indicating that the result
we have obtained in our lattice calculation should be corrected
upward by about $0.3$ MeV, or about 6\%. In upcoming studies
this estimate will be checked directly on larger box sizes.

\section{Vector Mesons}

\begin{table}[b]
\begin{center}
\caption[tbl:VChPT]{Coefficients (in lattice units) of fitting function, 
Eq.(\ref{eq:VChPT}) for vector mesons.}
\begin{tabular}{ll}
  & Fit \\
\hline
 $A$ & $ 0.567(1) + 0.0068(1) (e_q + e_{\bar q})^2 $ \\ 
 $B$ & $ 0.523(47) + 0.205(22) e_q^2 + 0.073(9) e_{\bar q}^2 $ \\
     & $ + 0.138(9) e_q e_{\bar q} - 0.027(1) e_q^2 e_{\bar q}^2 $ \\
     & $ - 0.017(1) e_q^4 - 0.013(1) e_{\bar q}^4 $ \\
     & $ - 0.009(1) e_q^3 e_{\bar q} -0.013(1) e_q e_{\bar q}^3$ \\ 
 $C$ & $ 0.81(29) $ \\
 $D$ & $ 0.38(50) $ \\
\end{tabular}
\end{center}
\end{table} 

The same techniques can be applied to vector mesons and
heavy-light mesons.
For the light vector mesons the expected form of the mass matrix includes
the following terms. 
\begin{eqnarray}
\label{eq:VChPT}
\lefteqn{m_{V} = A(e_q, e_{\bar{q}}) +  m_q B(e_q,e_{\bar{q}}) 
            + m_{\bar{q}}B(e_{\bar{q}},e_q)} \nonumber \\
   & & +(m_q^2+m_{\bar q}^2)C(e_q, e_{\bar{q}}) 
              +  m_q m_{\bar q}D(e_q,e_{\bar{q}})  
\end{eqnarray}
(Including nonanalytic terms{\cite{Xlogs} (e.g. $O(m^{3/2})$)
does not subtantially alter our final results.)
Calculating all the vector mesons mass combinations and fitting to this 
form, we obtain the coefficients shown in Table 2. 
From these results we obtain the mass differences 
\begin{eqnarray}
 m_{\rho^+} - m_{\rho^0}~~ &=& -0.74 \pm 1.14{(\rm stat)} {\rm ~MeV} \\
 m_{K^{*0}} - m_{K^{*+}} &=& 3.58 \pm 0.77{(\rm stat)}{\rm ~MeV} \;
\end{eqnarray}

\section{Conclusions}

Here we have focused mainly on the pseudoscalar meson masses. This is the 
most precise way of determining the quark masses as well as 
providing an important test of the method in  the $\pi^+-\pi^0$ splitting
\cite{emlatt}.
Further calculations of electromagnetic splittings in the vector mesons and
the baryons \cite{baryon}, as well as in heavy-light systems, are possible
using the present method. This will provide an extensive opportunity
to test the precision of the method and gain confidence in the results.
Eventually reliable lattice calculations
for all isospin breaking effects should be possible. 
 
We thank Tao Han, George Hockney, Paul Mackenzie and 
Tetsuya Onogi for contributions to our effort.




\begin{thebibliography}{99}
\bibitem{pdata}\underline{Review of Particle Properties}, Phys. Rev. {\bf D50}
(1994).
\bibitem{leutrev} J.~Gasser and H.~Leutwyler, Phys. Rep. 87 (1982) 77.
\bibitem{quark_mass} P.~B.~Mackenzie in these proceedings.
\bibitem{emlatt} A.~Duncan, E.~Eichten and H.~Thacker, Phys. Rev. Lett. 
{\bf 76} (1996) 3894.
\bibitem{etaprime} H.~Thacker in these proceedings.
\bibitem{Xlogs} S.~R.~Sharpe in these proceedings.
\bibitem{Gasser} J.~Gasser and H.~Leutwyler, Nucl. Phys. {\bf B250} (1985) 465.
\bibitem{1overN} W.~A.~Bardeen, J.~Bijnens and J.~M.~Gerard, Phys. Rev. Lett.
{\bf 62} (1989) 1343.
\bibitem{baryon} A.~Duncan in these proceedings.
\end{thebibliography}
\end{document}